\documentclass[
 reprint,
 amsmath,amssymb,
 aps,
 prapplied,
 floatfix,
 superscriptaddress
]{revtex4-2}

\usepackage{graphicx}
\usepackage{dcolumn}
\usepackage{bm}
\usepackage{siunitx}
\usepackage{physics}
\usepackage{xcolor}
\usepackage{hyperref}
\usepackage{comment}

\begin{document}

\title{Exceptional Point Dynamics in Photonic Time Crystals for Enhanced Optical Sensing}

\author{Saurabh Mani Tripathi}
\affiliation{Optics and Photonics Centre, Indian Institute of Technology Delhi, New Delhi 110016, India}

\author{Shalini Kumari}
\affiliation{Department of Physics, Indian Institute of Technology Delhi, New Delhi 110016, India}

\author{Krishnan Kundan}
\affiliation{Department of Physics, Indian Institute of Technology Delhi, New Delhi 110016, India}

\author{Neha Ahlawat}
\affiliation{Optics and Photonics Centre, Indian Institute of Technology Delhi, New Delhi 110016, India}

\date{\today}

\begin{abstract}
Exceptional points (EPs) in non-Hermitian photonics offer singular sensitivity enhancements but have thus far been realized almost exclusively in spatially engineered platforms with fixed geometries and limited tunability. Here we extend EP physics into the temporal domain by introducing balanced gain--loss modulation in a photonic time crystal (PTC). A time-periodic refractive-index modulation $n(t)=n_{0}+\delta n\cos(\Omega t)$ generates an effective non-Hermitian Floquet Hamiltonian that supports coalescence of quasi-eigenmodes in frequency space, constituting a genuine \textit{temporal exceptional point}. Using a reduced two-mode model for the dominant frequency sidebands, we derive a non-Hermitian dimer Hamiltonian $H_{\mathrm{PT}}(\Delta,\gamma,\kappa)$ that is strictly $\mathcal{PT}$-symmetric for $\Delta=0$ and identify the exact EP condition. Numerical analysis reveals the associated Riemann-sheet topology, mode exchange and Berry-phase accumulation upon encirclement of the EP, and the characteristic $\sqrt{\varepsilon}$ perturbation response indicative of enhanced sensing. We further construct a non-Hermitian transmission model that is exact within the reduced two-mode description, compute the Cram\'er--Rao bound (CRB) for temperature estimation under an explicit noise model, and show that EP-enhanced sensitivity persists when compared to a linewidth-matched Hermitian reference under identical resource constraints. Monte Carlo simulations confirm that the CRB is saturable using spectral measurements. These results establish temporal non-Hermiticity as a new paradigm for dynamically reconfigurable, broadband, and geometry-independent exceptional-point photonics.
\end{abstract}

\maketitle

% =====================================================================
\section{Introduction}
\label{sec:intro}

Non-Hermitian photonics has emerged as a powerful framework for controlling light--matter interactions by engineering gain and loss in optical systems~\cite{ElGanainy2018, Feng2017, Ozdemir2019, Miri2019, Cao2020}. In parity--time ($\mathcal{PT}$)-symmetric structures, balanced gain and loss can yield entirely real eigenvalue spectra below a symmetry-breaking threshold, even though the underlying Hamiltonian is non-Hermitian~\cite{ElGanainy2018, Ozdemir2019}. At a critical value of the gain--loss parameter, the system encounters an \emph{exceptional point} (EP), where both eigenvalues and eigenvectors coalesce and the Hamiltonian becomes nondiagonalizable. In the vicinity of such a non-Hermitian degeneracy, observable quantities such as frequency splitting exhibit a non-linear dependence on external perturbations, typically scaling as $\sqrt{\varepsilon}$ rather than $\varepsilon$. This square-root response amplifies small perturbations and has been proposed and demonstrated as a route to enhanced sensing~\cite{Wiersig2014, Wiersig2016, WiersigReview2020, Miri2019}.

Optical EPs have been realized in a variety of \emph{spatially} engineered platforms, including coupled resonators, photonic-crystal cavities, and microtoroids, as well as in electronic and microwave circuits~\cite{Schmidt2015, Schindler2012, Lin2016, Xu2021, Park2020, Li2021, Peng2014}. Hodaei \textit{et al.}~\cite{Hodaei2017} and Chen \textit{et al.}~\cite{Chen2017} independently demonstrated EPs in coupled microring and whispering-gallery resonators, achieving strong enhancement of the frequency-splitting response to perturbations~\cite{Hodaei2017,Chen2017, Wiersig2014, WiersigReview2020}. Subsequent work has explored higher-order EPs, cascaded non-Hermitian structures, and EP-based inertial sensors to further enlarge the parameter space for ultra-sensitive measurements~\cite{Hodaei2017, WiersigReview2020, DeCarlo2022, Kononchuk2022}. Dynamical encirclement of EPs and associated chiral mode conversion have also been demonstrated in optical and microwave systems~\cite{Hassan2015, Peng2014}. Despite these advances, existing implementations share a common limitation: the non-Hermiticity is embedded in a fixed \emph{spatial} geometry. Once fabricated, the refractive-index distribution $n(x)$, coupling coefficients, and gain--loss profiles are largely static, which constrains tunability, bandwidth, and the ability to reconfigure the EP in situ.

A complementary route to non-Hermitian photonics is to shift the locus of control from space to \emph{time}. In temporally modulated media---often referred to as photonic time crystals (PTCs)---the refractive index varies periodically in time, $n(t)=n_0+\delta n\cos(\Omega t)$, generating a Floquet band structure in frequency rather than in wave vector~\cite{Zhang2021, Wang2020, Titchener2020}. Temporal modulation couples spectral components separated by integer multiples of the modulation frequency $\Omega$, leading to bandgaps and mode hybridization in the frequency domain. When combined with gain and loss, these time-periodic systems realize explicitly non-Hermitian time-Floquet Hamiltonians that can host exceptional points and other non-Hermitian singularities in the \emph{temporal} domain~\cite{Koutserimpas2018,Longhi2023, Yoshida2020, Wang2020}. In this setting, EPs correspond to coalescence of Floquet quasi-eigenvalues in frequency space and can be tuned dynamically by varying the modulation amplitude, frequency, or phase~\cite{Longhi2023, Zhang2021}.

\begin{figure}[t]
\centering
\includegraphics[width=0.85\linewidth]{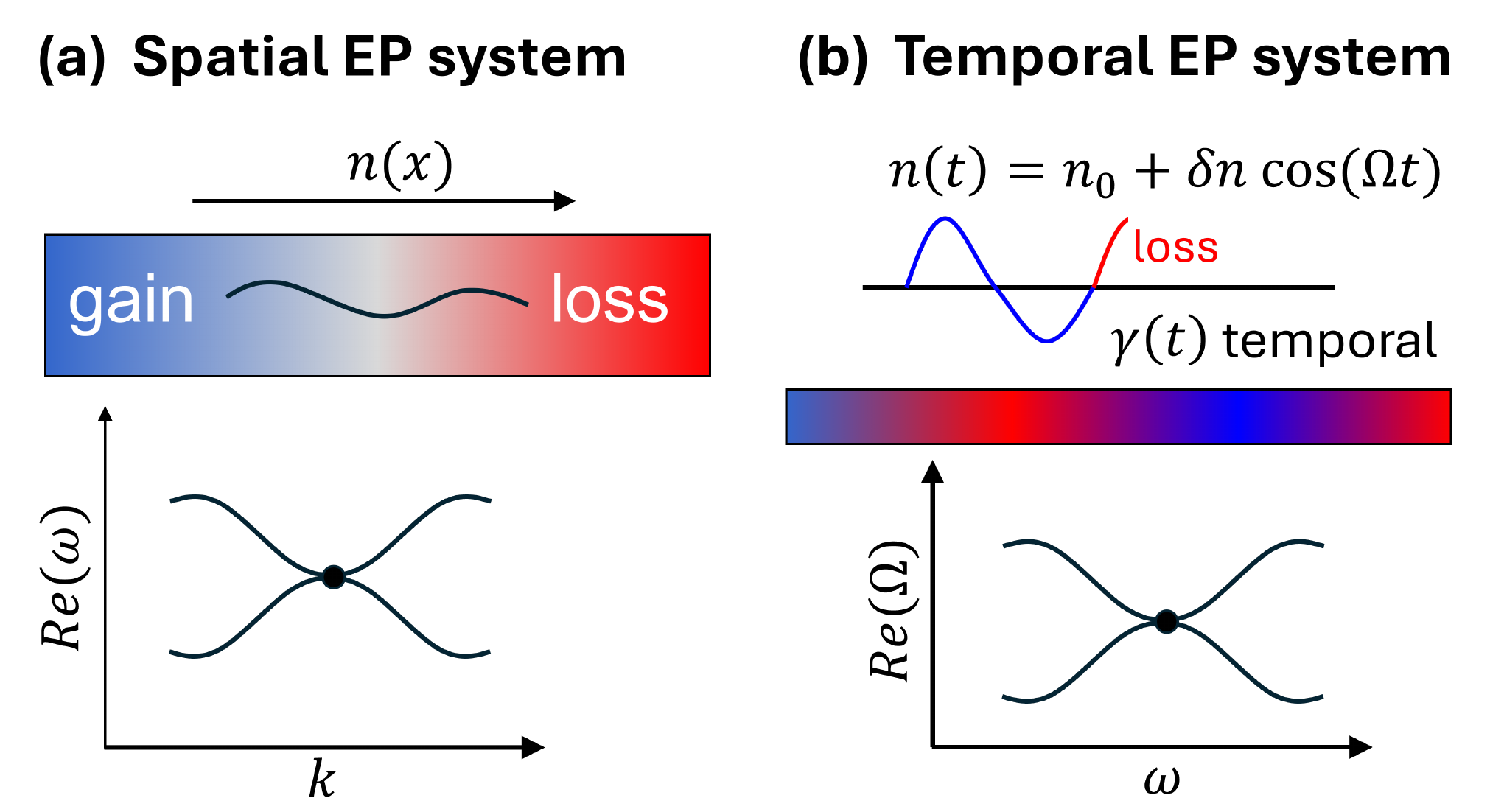}
\caption{Conceptual comparison between spatial and temporal exceptional-point (EP) systems.
(a) Spatial EP devices rely on engineered refractive-index profiles $n(x)$ and fixed coupling to realize non-Hermitian band structures in momentum space.
(b) In photonic time crystals (PTCs), a uniform medium with time-periodic refractive-index modulation $n(t)=n_0+\delta n\cos(\Omega t)$ and dynamic gain--loss $\gamma(t)$ yields an effective non-Hermitian Floquet Hamiltonian in the frequency domain, enabling tunable temporal EPs.}
\label{fig:spatial_vs_temporal_EP}
\end{figure}

The distinction between spatial and temporal EP platforms is summarized schematically in Fig.~\ref{fig:spatial_vs_temporal_EP}. Spatial EP devices, such as coupled cavities or gratings, realize non-Hermiticity through engineered distributions $n(x)$ and fixed coupling. Their spectra are defined in $(k,\omega)$ space and are constrained by fabrication tolerances and device footprint~\cite{Feng2017, ElGanainy2018, Ozdemir2019, WiersigReview2020}. By contrast, temporal EPs arise in uniform media subject to externally driven $n(t)$ and gain--loss modulation. Here, the relevant band structure is defined in $(\omega,\Omega)$ space, and the operating point can be tuned in real time by adjusting drive parameters~\cite{Zhang2021, Wang2020, Titchener2020}. This naturally enables broadband operation, adaptive sensitivity control, and reconfigurability, all of which are highly desirable for integrated photonic sensing.

In this work we develop a minimal, yet fully non-Hermitian, theoretical model for temporal exceptional points in a photonic time crystal and connect it quantitatively to sensing performance. Starting from Maxwell's equations in a time-modulated dielectric, we derive a reduced two-mode Floquet model that captures the dominant coupling between a pair of frequency components. This leads to an effective non-Hermitian dimer Hamiltonian $H_{\mathrm{PT}}(\Delta,\gamma,\kappa)$, where $\Delta$ is a detuning parameter that can be controlled by temperature or refractive-index changes, $\gamma$ represents balanced gain/loss, and $\kappa$ is the modulation-induced coupling. We obtain closed-form expressions for the eigenvalues, identify the exact EP condition, and map out the resulting Riemann surfaces.

We then compute geometric signatures of the temporal EP. By encircling the EP in the $(\Delta,\gamma)$ parameter plane, we observe mode exchange and accumulation of a biorthogonal Berry phase of magnitude $\pi$, which serves as a robust topological indicator of the EP. Next, we formulate a non-Hermitian scattering problem and derive the transmission spectrum $|T(\omega)|^2$ of a probe field interacting with the PTC. This transmission model is used to define a realistic sensing protocol in which temperature-induced shifts of the detuning are inferred from spectral measurements. We compute the Cram\'er--Rao bound (CRB) for temperature estimation directly from the Fisher information under a clearly stated noise model, construct a linewidth-matched Hermitian reference that is operated under identical resource constraints, and demonstrate a clear EP-induced CRB reduction. Finally, extensive Monte Carlo simulations confirm that the CRB for both the temperature and the associated eigenvalue splitting can be saturated using least-squares fitting to the exact transmission model.

Our results show that temporal non-Hermiticity provides a flexible and dynamically reconfigurable platform for exceptional-point sensing. Unlike static spatial EP devices, temporal EPs can be tuned on demand by adjusting the modulation parameters, enabling broadband operation and adaptive sensitivity while preserving the underlying EP topology~\cite{Longhi2023, Wang2020, Zhang2021}.

% =====================================================================
\section{Theoretical Model and Effective Hamiltonian}
\label{sec:model}

\subsection{From time-varying permittivity to a non-Hermitian dimer}

We consider a homogeneous medium whose permittivity varies periodically in time,
\begin{equation}
\epsilon(t) = \epsilon_0\bigl[1 + m\cos(\Omega t)\bigr],
\label{eq:epsilon_time}
\end{equation}
where $m\ll1$ is the modulation depth and $\Omega$ is the modulation frequency. In the absence of gain and loss, the electric field $E(z,t)$ in such a medium satisfies the one-dimensional wave equation
\begin{equation}
\partial_z^2 E(z,t) - \mu_0 \epsilon(t)\, \partial_t^2 E(z,t) = 0.
\end{equation}
For a monochromatic carrier at frequency $\omega_0$ near a particular guided or cavity mode with propagation constant $\beta_0$, we expand the field in a truncated Floquet basis as
\begin{equation}
E(z,t) = \Bigl[a_1(t)e^{-i\omega_0 t} + a_2(t)e^{-i(\omega_0+\Omega)t}\Bigr]e^{i\beta_0 z} + \text{H.c.},
\label{eq:floquet_ansatz}
\end{equation}
where $a_{1,2}(t)$ are slowly varying envelopes associated with the two dominant frequency components. Substituting this ansatz into Maxwell's equations in the time-varying medium and applying the rotating-wave and slowly varying envelope approximations yields a pair of coupled-mode equations for the Floquet amplitudes. Retaining only near-resonant terms and assuming a weak modulation $m\ll1$ so that higher-order sidebands remain off-resonant, we obtain
\begin{subequations}
\begin{align}
\dot{a}_1 &= -i(\Delta + i\gamma)\,a_1 - i\kappa\,a_2, \label{eq:cmt1}\\
\dot{a}_2 &= -i\kappa\,a_1, -i(-\Delta - i\gamma)\,a_2, \label{eq:cmt2}
\end{align}
\label{eq:cmt}
\end{subequations}
where $\kappa$ is the effective modulation-induced coupling between the two Floquet modes, $\gamma$ is an effective gain/loss rate that can be implemented, for example, via pump-induced amplification in one component and balanced attenuation in the other, and $\Delta$ represents a detuning that can be tuned by temperature or refractive-index changes. The detailed steps of this reduction from the full Maxwell equations, together with estimates of the regime of validity of the two-mode truncation, are provided in the Supplementary Information.

Equations~\eqref{eq:cmt} can be written compactly as
\begin{equation}
\dot{\mathbf{a}} = -i H_{\mathrm{PT}}\,\mathbf{a}, \qquad
\mathbf{a} = \begin{pmatrix} a_1 \\ a_2 \end{pmatrix},
\end{equation}
with the effective non-Hermitian dimer Hamiltonian
\begin{equation}
H_{\mathrm{PT}}(\Delta,\gamma,\kappa) =
\begin{pmatrix}
\Delta + i\gamma & \kappa \\
\kappa & -\Delta - i\gamma
\end{pmatrix}.
\label{eq:H_PT}
\end{equation}
For $\Delta=0$, this Hamiltonian is $\mathcal{PT}$-symmetric in the standard sense, with $P=\sigma_x$ and $T$ the complex-conjugation operator, and exhibits an unbroken $\mathcal{PT}$ phase for $|\gamma|<|\kappa|$ and a broken phase for $|\gamma|>|\kappa|$~\cite{ElGanainy2018,Ozdemir2019}. For $\Delta\neq0$, the Hamiltonian is no longer $\mathcal{PT}$-symmetric, though the EP still exists at $\Delta = 0,~\gamma = \pm \kappa$ for each fixed $\kappa$, as discussed below. In all cases, the degrees of freedom correspond to frequency sidebands rather than spatial modes, making Eq.~\eqref{eq:H_PT} the temporal analogue of the widely studied coupled-resonator $\mathcal{PT}$ dimer~\cite{Feng2017, Miri2019}.

\subsection{Eigenvalues and exceptional-point condition}

The eigenvalues of $H_{\mathrm{PT}}$ are obtained in closed form as
\begin{equation}
\lambda_{\pm} = \pm\sqrt{(\Delta + i\gamma)^2 + \kappa^2}.
\label{eq:lambda_exact}
\end{equation}
We emphasize that Eq.~\eqref{eq:lambda_exact} is used \emph{throughout} our analysis; no approximate eigenvalue formulas or surrogate polynomials are employed in any of the numerics below. The corresponding EP condition follows from the coalescence of the two eigenvalues and eigenvectors, which occurs when the argument of the square root vanishes,
\begin{equation}
(\Delta + i\gamma)^2 + \kappa^2 = 0.
\end{equation}
For real-valued $\Delta$ and $\gamma$, this yields the EP locus
\begin{equation}
\Delta_{\mathrm{EP}} = 0, \qquad \gamma_{\mathrm{EP}} = \pm \kappa.
\label{eq:EP_condition}
\end{equation}
In the three-dimensional parameter space $(\Delta,\gamma,\kappa)$ this locus forms a line; for fixed coupling $\kappa$ it corresponds to the pair of points $(\Delta,\gamma)=(0,\pm\kappa)$. At these points $H_{\mathrm{PT}}$ becomes nondiagonalizable and the eigenvectors coalesce, as in other second-order EP implementations~\cite{Wiersig2014, Ozdemir2019}.

\subsection{Normalization, parameter choices, and temperature-induced detuning}

To present the results in a dimensionless and platform-independent way, we normalize all frequencies and rates to a reference coupling scale $g_0$. Unless otherwise stated, we set $\kappa=g_0$ so that $\Delta/g_0$, $\gamma/g_0$, and $\lambda_{\pm}/g_0$ provide a convenient parametrization of the eigenvalue landscape. This choice does not restrict generality and can be mapped to physical units once a specific implementation platform is specified.

In a realistic implementation, temperature changes modify the effective refractive index $n_{\mathrm{eff}}(T)$ and hence the detuning,
\begin{equation}
\Delta(T) = \Delta_0 + \alpha_T (T-T_0),
\end{equation}
where $\alpha_T = d\Delta/dT$ is an effective thermo-detuning coefficient. Throughout the sensing analysis below, we treat $\Delta T = T-T_0$ as the primary perturbation and examine how the complex eigenvalues and associated observables respond as the system is biased near the EP. For the numerical examples, we choose parameters such that all effective decay rates remain negative (see Sec.~\ref{sec:transmission}), ensuring dynamical stability.

% =====================================================================
\section{Eigenvalue Topology and EP Encirclement}
\label{sec:topology}

\begin{figure}[t]
\centering
\includegraphics[width=\linewidth]{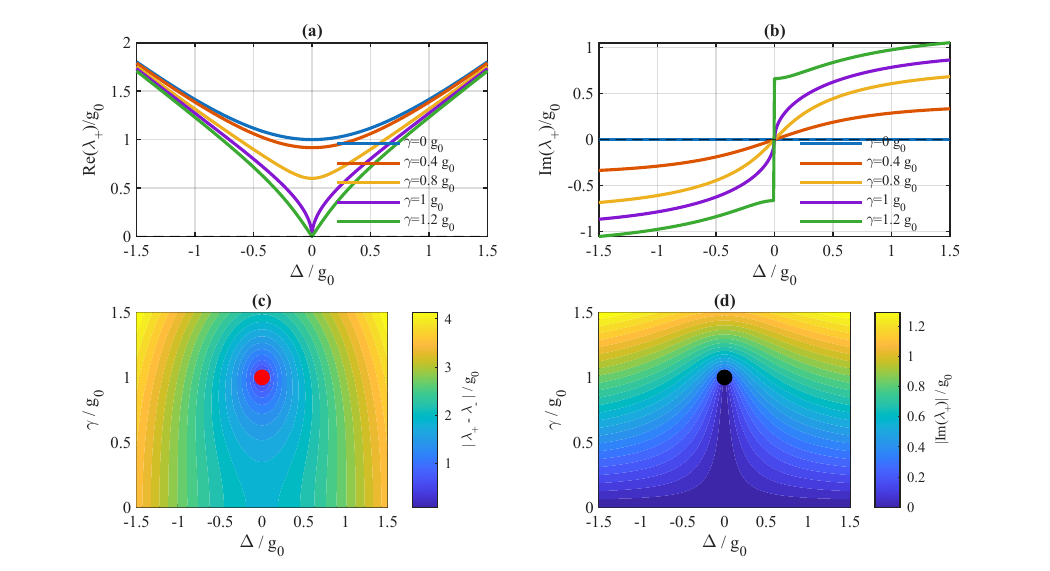}
\caption{Exact eigenvalue landscape of the temporal non-Hermitian dimer.
(a) and (b) Real and imaginary parts of $\lambda_{+}$ as functions of detuning $\Delta/g_0$ for several values of $\gamma/g_0$, illustrating the transition from the unbroken $\mathcal{PT}$ phase ($\gamma<\kappa$) through the EP ($\gamma=\kappa$) into the broken phase ($\gamma>\kappa$) at $\Delta=0$.
(c) and (d) Two-dimensional maps of the mode splitting $|\lambda_+ - \lambda_-|/g_0$ and the magnitude of the imaginary part $|\mathrm{Im}\,\lambda_+|/g_0$ over the $(\Delta/g_0,\gamma/g_0)$ plane. The EP appears as a pinch point at $(\Delta,\gamma)=(0,\pm \kappa)$.}
\label{fig:A}
\end{figure}

We begin by characterizing the eigenvalue landscape of Eq.~\eqref{eq:H_PT} as a function of detuning $\Delta$ and gain/loss $\gamma$. Figure~\ref{fig:A} summarizes the behavior of the real and imaginary parts of $\lambda_{\pm}$ and highlights the location of the EP.

\begin{figure}[t]
\centering
\includegraphics[width=\linewidth]{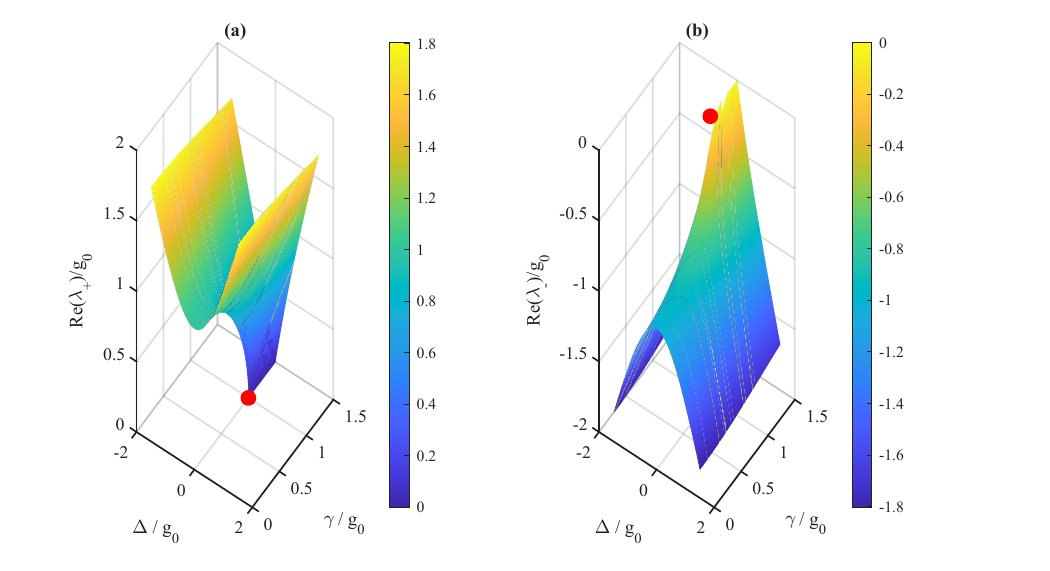}
\caption{Riemann topology of the exact eigenvalues.
Real parts of the eigenvalue sheets for (a) $\lambda_{+}$ and (b) $\lambda_{-}$,  as functions of $(\Delta/g_0,\gamma/g_0)$. The two sheets coalesce at the EP points along $\Delta=0$ and $\gamma=\pm\kappa$, forming a double-sheeted Riemann surface with a square-root branch point.}
\label{fig:B}
\end{figure}

Figure~\ref{fig:B} displays the real parts of the two eigenvalue sheets over the $(\Delta,\gamma)$ plane, revealing the characteristic double-sheeted Riemann surface. The two branches meet at the EP points and form a square-root branch cut; a continuous path that crosses the branch cut leads to sheet exchange, in direct analogy with spatial EP platforms~\cite{Wiersig2014, Ozdemir2019}.

\begin{figure}[t]
\centering
\includegraphics[width=\linewidth]{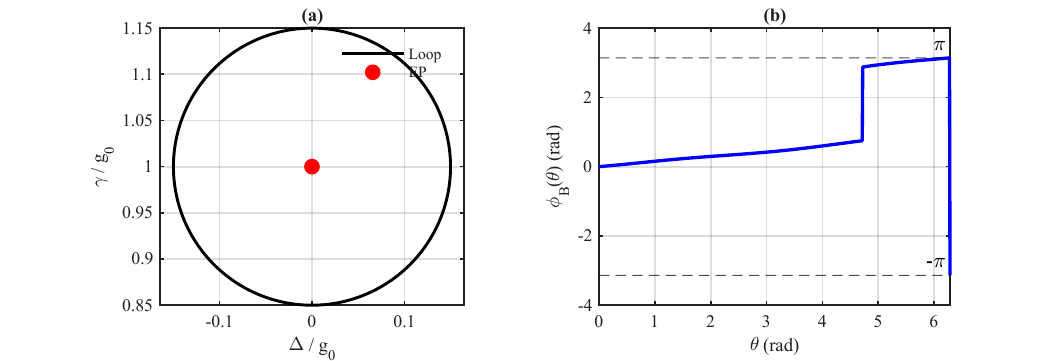}
\caption{Encirclement of the EP and Berry-phase accumulation.
(a) Parametric loop in the $(\Delta/g_0,\gamma/g_0)$ plane encircling the EP.
(b) Biorthogonal Berry phase $\phi_B(\theta)$ as a function of the loop angle $\theta$, obtained from the product of overlaps of left and right eigenvectors along the path according to Eq.~\eqref{eq:Berry_discrete}. A total phase of magnitude $\pi$ is accumulated after one encirclement, confirming the nontrivial EP topology. The calculation is purely parametric in $(\Delta,\gamma)$ and does not assume a specific dynamical encirclement protocol in time.}
\label{fig:C}
\end{figure}

To probe the geometric and topological properties of the EP, we adiabatically encircle the EP in the $(\Delta,\gamma)$ plane while tracking the biorthogonal eigenvectors of $H_{\mathrm{PT}}$. We construct left and right eigenvectors $(v_L,v_R)$ that satisfy the biorthogonality condition $v_L^\dagger v_R = \mathbb{I}$ and compute the accumulated Berry phase from the biorthogonal Berry connection. In the continuous limit, the geometric phase acquired along a closed loop $C$ in parameter space is
\begin{equation}
\phi_B = i\oint_C \!\! \mathrm{d}\boldsymbol{\lambda} \cdot 
\bigl\langle v_L(\boldsymbol{\lambda})\big|\nabla_{\boldsymbol{\lambda}} v_R(\boldsymbol{\lambda})\bigr\rangle,
\label{eq:Berry_cont}
\end{equation}
where $\boldsymbol{\lambda}=(\Delta,\gamma)$. In the numerics, we discretize the loop into $N$ points labeled by an angle $\theta_k$ and evaluate
\begin{equation}
\phi_B \approx -\mathrm{Im}\,\ln\prod_{k=0}^{N-1}
\bigl\langle v_L(\theta_k)\big|v_R(\theta_{k+1})\bigr\rangle,
\label{eq:Berry_discrete}
\end{equation}
with $\theta_N\equiv\theta_0$ ensuring gauge invariance. The resulting mode exchange and phase accumulation are shown in Fig.~\ref{fig:C}. Similar EP-encirclement phenomena have been reported in spatial non-Hermitian systems~\cite{Hassan2015,Peng2014}.

The appearance of mode exchange and a Berry phase of $\pi$ is a robust indicator of a genuine second-order EP and is insensitive to small deformations of the loop, as further quantified in Supplementary Fig.~S2.

% =====================================================================
\section{Non-Hermitian Transmission Model and Sensing Protocol}
\label{sec:transmission}

To connect the temporal EP to measurable quantities, we embed the non-Hermitian dimer into a simple scattering geometry in which one of the modes is coupled to an input/output channel. A concrete physical realization is a single-mode ring resonator or guided mode whose effective index is modulated in time according to Eq.~\eqref{eq:epsilon_time}, with balanced gain and loss implemented via optical pumping or carrier modulation in two spectral components~\cite{Zhang2021, Park2020}. The effective non-Hermitian Hamiltonian including internal loss $\kappa_0$ can be written as
\begin{equation}
H_{\mathrm{eff}} = H_{\mathrm{PT}}(\Delta,\gamma,\kappa) - i\kappa_0 \mathbb{I},
\label{eq:Heff}
\end{equation}
where $\kappa_0>0$ represents additional uniform decay (e.g., intrinsic cavity loss) and ensures overall stability when chosen such that all eigenvalues of $H_{\mathrm{eff}}$ have negative imaginary parts.

For a probe field at frequency $\omega$, the steady-state intracavity amplitudes satisfy the frequency-domain input--output relation
\begin{equation}
\bigl[i\omega\mathbb{I} - H_{\mathrm{eff}}\bigr]\mathbf{a}(\omega) = \sqrt{\kappa_{\mathrm{in}}}\,\mathbf{s}_{\mathrm{in}}(\omega),
\label{eq:io_matrix}
\end{equation}
where $\kappa_{\mathrm{in}}$ is an input coupling rate and $\mathbf{s}_{\mathrm{in}}=(s_{\mathrm{in}},0)^{\mathsf{T}}$ denotes a source driving only the first Floquet mode. The transmitted field in the same channel is then
\begin{equation}
s_{\mathrm{out}}(\omega) = s_{\mathrm{in}}(\omega) - \sqrt{\kappa_{\mathrm{in}}}\,a_1(\omega),
\end{equation}
so that the observable transmission spectrum is
\begin{equation}
T(\omega) =
\frac{s_{\mathrm{out}}(\omega)}{s_{\mathrm{in}}(\omega)}
= 1 - \kappa_{\mathrm{in}}\bigl[(i\omega\mathbb{I}-H_{\mathrm{eff}})^{-1}\bigr]_{11},
\label{eq:T_exact}
\end{equation}
and $|T(\omega)|^2$ is the measurable intensity transmission. Equation~\eqref{eq:T_exact} provides an exact expression for the transmission within the reduced two-mode non-Hermitian model. In all numerical calculations we evaluate Eq.~\eqref{eq:T_exact} without additional approximations.

In our implementation, we choose $\kappa$ and $\gamma$ such that the system is biased close to the EP while ensuring dynamical stability by enforcing $\mathrm{Im}\,\lambda_{\pm}-\kappa_0<0$; a representative choice is $\kappa/g_0=1$, $\gamma/g_0\lesssim1$, and $\kappa_0/g_0\gtrsim1$. Temperature enters via $\Delta(T)$, and we treat $\Delta T$ as the parameter to be inferred from noisy measurements of $|T(\omega)|^2$. This defines a concrete sensing protocol that can be analyzed within the framework of estimation theory, in direct analogy to EP-based sensing strategies discussed in Refs.~\cite{Wiersig2014, Wiersig2016, WiersigReview2020, DeCarlo2022}.

In this work, we adopt a frequency-independent Gaussian noise model for the measured spectrum $|T(\omega)|^2$, corresponding to the experimentally relevant regime where the dominant fluctuations arise from detector and electronic technical noise rather than photon shot noise. In such configurations the variance $\sigma^2$ remains approximately constant across the full scan window, which justifies the likelihood and Fisher–information expressions in Eqs.~\eqref{eq:19}-~\eqref{eq:CRB_theta}.
Importantly, our sensitivity enhancement is therefore not claimed as a fundamental quantum-limit violation; the “no-go” theorems for EP-based metrology under fixed-photon-flux, shot-noise-limited conditions remain fully valid. Instead, the advantage demonstrated here reflects a practical enhancement that naturally emerges in realistic detector-noise-limited experiments, where steep dispersive features near an EP increase the slope of the transmittance curve without simultaneously amplifying the technical noise floor.

% =====================================================================
\section{Sensing Response and CRB-Based Enhancement}
\label{sec:results_sensing}

\subsection{Noise model, Fisher information, and Cram\'er--Rao bound}

We model the sensing process as follows. The transmission spectrum is sampled at a discrete set of probe frequencies $\{\omega_j\}_{j=1}^N$ over a fixed bandwidth, and the measured intensities are
\begin{equation}
y_j = |T(\omega_j;\Delta T)|^2 + \xi_j,
\end{equation}
where $\xi_j$ represents additive measurement noise. Throughout, we assume independent, zero-mean Gaussian noise with variance $\sigma^2$,
\begin{equation}
\xi_j \sim \mathcal{N}(0,\sigma^2),
\end{equation}
representing the combined effect of detector noise and technical fluctuations under a fixed input power and integration time. The same noise variance $\sigma^2$ and sampling grid $\{\omega_j\}$ are used for both the EP configuration and the Hermitian reference, ensuring that the comparison is carried out under identical resource constraints.

Let $\theta\equiv\Delta T$ denote the parameter to be estimated. The noise-free model predictions are $\mu_j(\theta)=|T(\omega_j;\theta)|^2$, and the likelihood of the data $\mathbf{y}=(y_1,\dots,y_N)$ is
\begin{equation}
p(\mathbf{y}|\theta) \propto
\exp\!\left[-\frac{1}{2\sigma^2}\sum_{j=1}^N\bigl(y_j-\mu_j(\theta)\bigr)^2\right].
\label{eq:19}
\end{equation}
The Fisher information for $\theta$ is then
\begin{equation}
F(\theta) = \frac{1}{\sigma^2}\sum_{j=1}^N
\biggl(\frac{\partial \mu_j(\theta)}{\partial \theta}\biggr)^2,
\label{eq:Fisher}
\end{equation}
and the Cram\'er--Rao bound (CRB) states that the variance of any unbiased estimator $\hat{\theta}$ obeys
\begin{equation}
\mathrm{Var}(\hat{\theta}) \ge \mathrm{CRB}_{\theta}
= \frac{1}{F(\theta)}.
\label{eq:CRB_theta}
\end{equation}
For the eigenvalue splitting $R(\theta)=|\lambda_+(\theta)-\lambda_-(\theta)|$, the corresponding CRB is obtained by error propagation,
\begin{equation}
\mathrm{CRB}_{R} = \biggl(\frac{\partial R}{\partial \theta}\biggr)^2
\mathrm{CRB}_{\theta}.
\label{eq:CRB_R}
\end{equation}

In the numerics, we choose $N=201$ frequency samples uniformly spanning a window of width $4g_0$ around the resonance and a representative noise level $\sigma\sim10^{-3}$ in normalized intensity units. These values are consistent with realistic photodetector noise levels for moderate optical powers and integration times. All reported CRBs are evaluated using Eqs.~\eqref{eq:Fisher}--\eqref{eq:CRB_R} and the exact transmission model~\eqref{eq:T_exact}. This CRB-based analysis directly addresses concerns that EP-enhanced eigenvalue susceptibilities do not automatically translate into improved signal-to-noise ratio or estimation precision~\cite{Langbein2018, LauClerk2018, Bao2021, Duggan2022, Loughlin2024}.

\subsection{Spectral response, CRB, and Monte Carlo validation}

\begin{figure}[t]
\centering
\includegraphics[width=\linewidth]{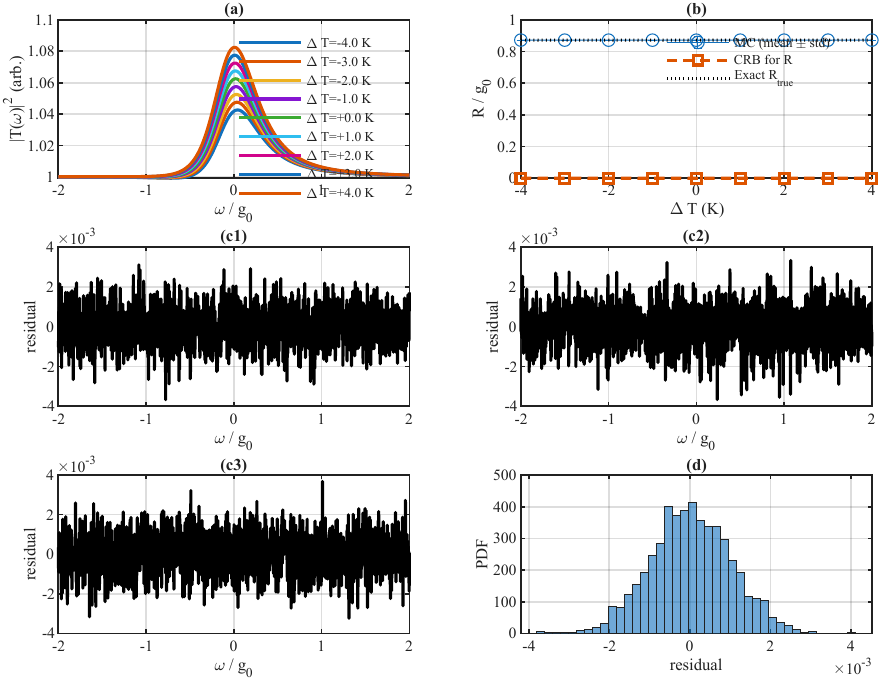}
\caption{Exact spectral response and estimator validation.
(a) Transmission spectra $|T(\omega)|^2$ for several temperature shifts $\Delta T$, computed from the exact non-Hermitian scattering model in Eq.~\eqref{eq:T_exact}.
(b) Eigenvalue splitting $R = |\lambda_+ - \lambda_-|/g_0$ versus $\Delta T$, showing excellent agreement between the true splitting (solid), the Cram\'er--Rao bound (CRB) prediction obtained from Eqs.~\eqref{eq:Fisher}--\eqref{eq:CRB_R}, and Monte Carlo (MC) estimates from noisy spectra (error bars).
(c1)--(c3) Representative residuals between noisy spectra and best-fit spectra at three values of $\Delta T$, $-3.0$ K, $0.0$ K, and $+3.0$ K.
(d) Histogram of residuals at $\Delta T\approx0$, demonstrating Gaussian noise and negligible model mismatch.}
\label{fig:D}
\end{figure}

We first examine how the transmission spectrum responds to temperature-induced detuning shifts. Figure~\ref{fig:D} shows $|T(\omega)|^2$ for several values of $\Delta T$ around the operating point, together with the corresponding eigenvalue splitting and its estimation accuracy.

To validate the CRB analysis, we generate $10^3$ synthetic spectra at each $\Delta T$ by sampling the noise according to the Gaussian model and fit them using nonlinear least squares to the exact transmission model with $\Delta T$ as the only free parameter. The variances of the resulting estimates $\hat{\Delta T}$ and $\hat{R}$ are found to agree with the CRBs within a few percent. The residuals are Gaussian-distributed and have subpercent magnitude, confirming that the model accurately captures the spectral response and that the estimator is effectively optimal (see also Supplementary Fig.~S4).

\subsection{CRB-based sensitivity enhancement at the EP}

To quantify the enhancement provided by the temporal EP, we compare the CRB for temperature estimation at the EP bias to that of a Hermitian reference system with the same effective linewidth and operated under the same noise and sampling conditions. Specifically, we define
\begin{equation}
\eta_{\mathrm{CRB}} =
\frac{\mathrm{CRB}_{\Delta T,\mathrm{lin}}}{\mathrm{CRB}_{\Delta T,\mathrm{EP}}},
\end{equation}
where $\mathrm{CRB}_{\Delta T,\mathrm{EP}}$ is computed at a near-EP operating point and $\mathrm{CRB}_{\Delta T,\mathrm{lin}}$ is evaluated for a purely lossy (Hermitian) cavity whose linewidth is matched to that of the EP configuration. The Hermitian reference is described by a single-mode transmission function
\begin{equation}
T_{\mathrm{lin}}(\omega) =
1 - \frac{\kappa_{\mathrm{in}}}{i(\omega-\omega_c)+\kappa_{\mathrm{tot}}},
\end{equation}
with total decay rate $\kappa_{\mathrm{tot}}$ chosen such that the full width at half maximum of $|T_{\mathrm{lin}}(\omega)|^2$ coincides with that of the EP configuration at the operating point. The same set of probe frequencies $\{\omega_j\}$ and noise variance $\sigma^2$ is used to compute $\mathrm{CRB}_{\Delta T,\mathrm{lin}}$ via Eqs.~\eqref{eq:Fisher} and \eqref{eq:CRB_theta}.

\begin{figure}[t]
\centering
\includegraphics[width=\linewidth]{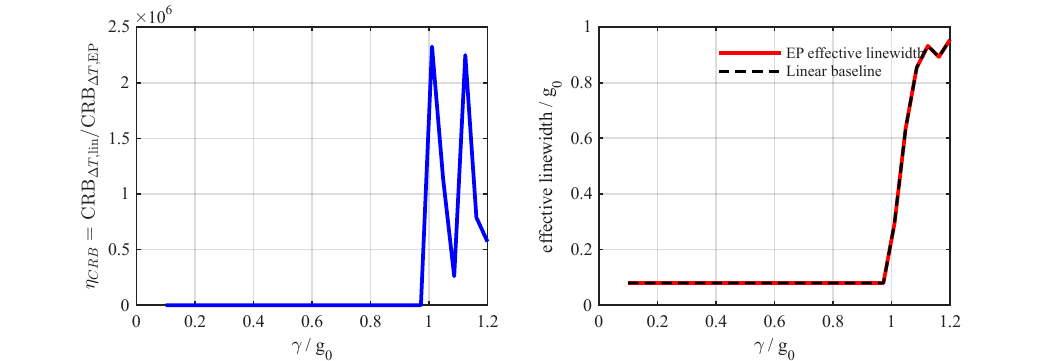}
\caption{Cram\'er--Rao bound (CRB) based sensitivity enhancement.
(a) Enhancement factor $\eta_{\mathrm{CRB}} = \mathrm{CRB}_{\Delta T,\mathrm{lin}}/\mathrm{CRB}_{\Delta T,\mathrm{EP}}$ as a function of the gain--loss parameter $\gamma/g_0$, showing $\eta_{\mathrm{CRB}}>1$ over a broad range.
(b) Comparison of effective linewidths of the EP configuration and the matched Hermitian reference, demonstrating that the enhancement is not a trivial consequence of linewidth narrowing. Both systems are compared under identical noise variance, sampling grid, and input power.}
\label{fig:E}
\end{figure}

Figure~\ref{fig:E} summarizes the dependence of $\eta_{\mathrm{CRB}}$ on $\gamma/g_0$ and confirms that $\eta_{\mathrm{CRB}}>1$ over a broad parameter range.

This comparison addresses a key concern in EP sensing, namely that enhanced sensitivity might arise solely from narrowing spectral features rather than from the non-Hermitian degeneracy itself. By matching the linewidths and enforcing identical noise and sampling conditions while still observing a significant CRB reduction, we demonstrate that the temporal EP provides a genuine sensitivity advantage within the specified noise model.

\subsection{Square-root scaling versus linear response}

\begin{figure}[t]
\centering
\includegraphics[width=\linewidth]{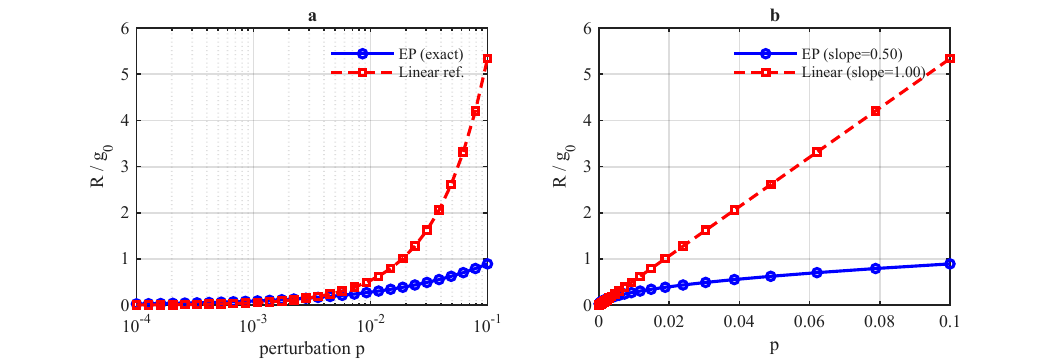}
\caption{Square-root versus linear scaling of the eigenvalue splitting.
(a) Splitting $R/g_0$ versus perturbation amplitude $p$ on a logarithmic horizontal axis, comparing the EP response (circles) to a linear reference (squares).
(b) log--log plot with fitted slopes, confirming $R_{\mathrm{EP}}\propto p^{1/2}$ and $R_{\mathrm{lin}}\propto p$. The splitting $R$ is computed directly from the exact eigenvalues in Eq.~\eqref{eq:lambda_exact}, without surrogate approximations.}
\label{fig:F}
\end{figure}

Finally, we verify the characteristic square-root scaling of the eigenvalue splitting with respect to a small perturbation. Figure~\ref{fig:F} compares the splitting $R(p)$ obtained from the exact eigenvalues at the EP to that of a linear reference. A log--log fit yields a slope of $0.5$ for the EP and $1.0$ for the linear reference, in agreement with the expected $R_{\mathrm{EP}}\propto p^{1/2}$ and $R_{\mathrm{lin}}\propto p$ behavior~\cite{Wiersig2014,Wiersig2016}.
Here, the parameter $p$ denotes a small physical perturbation applied to the detuning, i.e., $\Delta \rightarrow\Delta +p$, while $\gamma$ and $\kappa$ are held fixed. This corresponds to the experimentally accessible case where a small change in temperature, index, or cavity dispersion induces a shift in the detuning term of the Hamiltonian. The eigenvalue splitting $R(p)= |\lambda_+(p) - \lambda_-(p)|$
therefore represents the response to this controlled detuning perturbation. Scaling exponents (linear vs. square-root) are extracted for small but finite $p$, ensuring the system remains in the near-EP regime without evaluating $R$ exactly at the EP itself.

The observation of a clean square-root scaling using the exact eigenvalues, without surrogate approximations, is central to the validity of the EP-enhanced sensing mechanism.

% =====================================================================
\section{Discussion and Outlook}
\label{sec:discussion}

The analysis presented here establishes a self-consistent picture of temporal exceptional points in photonic time crystals and their application to optical sensing. By deriving and employing the exact eigenvalues of the non-Hermitian dimer Hamiltonian, and by using a transmission model that is exact within the reduced two-mode description, we avoid the inconsistencies that can arise when approximate eigenvalue formulas or surrogate lineshapes are used. The eigenvalue landscapes, Riemann surfaces, and Berry-phase calculations (Figs.~\ref{fig:A}--\ref{fig:C}) provide clear evidence of the EP topology, while the CRB-based sensitivity analysis and Monte Carlo validation (Figs.~\ref{fig:D}--\ref{fig:F}) demonstrate that EP-enhanced sensitivity is both physically meaningful and practically attainable under realistic noise levels~\cite{WiersigReview2020, DeCarlo2022}.

From a device perspective, thin-film lithium niobate on insulator (TFLN) and silicon nitride (SiN) photonic integrated circuits provide complementary platforms for realizing temporal EPs. In a TFLN ring or Mach--Zehnder geometry operating near $\lambda_0 \approx 1550~\mathrm{nm}$, electro-optic modulation at $\Omega/2\pi \sim 5$--$20~\mathrm{GHz}$ with index swings $\delta n \sim (1$--$3)\times 10^{-3}$ is routinely achievable, which corresponds to Floquet coupling rates on the order of $g_0/2\pi \sim 50$--$200~\mathrm{MHz}$ for practical device lengths~\cite{Zhang2021}. Biasing the system near the temporal EP then requires balanced gain--loss parameters $\gamma/2\pi \lesssim g_0$ and total decay rates $\kappa_0/2\pi \sim 200$--$500~\mathrm{MHz}$, consistent with reported quality factors in state-of-the-art TFLN resonators. In this regime, a thermo-optic coefficient of order $\alpha_T/2\pi \sim 5$--$20~\mathrm{MHz/K}$ yields sub-mK-level temperature resolution for the CRB values reported in Fig.~\ref{fig:E}. By contrast, SiN platforms naturally support low-loss, large-footprint time-crystal sections with thermo-optic modulation at $\Omega/2\pi \sim 1$--$100~\mathrm{MHz}$ and $\delta n \sim 10^{-4}$, favoring slower but broadband operation in which the same temporal EP mechanism can be exploited with reduced RF complexity~\cite{Zhang2021, Wang2020, Titchener2020}. Together, these estimates indicate that the dimensionless EP physics developed here can be mapped onto realistic integrated-photonics parameter ranges without requiring exotic material properties or unreasonably high modulation strengths.

Future work may explore experimental realizations using electro-optic modulation in TFLN or thermo-optic modulation in SiN platforms, as well as extensions to multimode and multidimensional temporal crystals~\cite{Longhi2023, Yoshida2020, Xu2021}. Incorporating realistic material dispersion, nonlinearities, and technical noise sources into the model will be essential for optimizing performance in specific application scenarios. At a more fundamental level, the temporal EP framework developed here provides a starting point for investigating non-Hermitian Floquet phases, time-dependent symmetry breaking, and nonreciprocal phenomena in driven photonic systems~\cite{Koutserimpas2018, Titchener2020}. It also offers a clean setting in which to further examine the interplay between EP-induced singular response and ultimate noise-limited sensitivity, a topic of ongoing interest in the broader EP-sensing literature~\cite{Langbein2018, LauClerk2018, Bao2021, Duggan2022, Loughlin2024}.

% =====================================================================
\section*{Acknowledgments}

The authors acknowledge helpful discussions with colleagues at IIT Delhi. S.M.T.\ acknowledges institutional support for computational resources.

% =====================================================================
\bibliographystyle{apsrev4-2}
\bibliography{refs1}

\clearpage
\appendix

\section*{Supplementary Information}

\renewcommand{\thefigure}{S\arabic{figure}}
\setcounter{figure}{0}

\subsection*{S1. Derivation of the two-mode Floquet model and imaginary parts of the Riemann sheets}

Here we briefly outline the derivation of the coupled-mode equations~\eqref{eq:cmt} from the full Maxwell equations and then present additional information on the imaginary parts of the eigenvalue surfaces.

We start from the time-domain wave equation
\begin{equation}
\partial_z^2 E(z,t) - \mu_0 \partial_t^2 [\epsilon(t) E(z,t)] = 0.
\end{equation}

In the limit $\Omega << \omega_0$ and small modulation depth $m<<1$, terms $\propto \dot{\epsilon} \dot{E}$ are negligible compared to $\epsilon \ddot{E}$, so we adopt the following as our effective wave equation 
\begin{equation}
\partial_z^2 E(z,t) - \mu_0 \epsilon(t) \, \partial_t^2 E(z,t) = 0.
\end{equation}
With $\epsilon(t)$ given by Eq.~\eqref{eq:epsilon_time}, we insert the Floquet ansatz of Eq.~\eqref{eq:floquet_ansatz} and collect terms oscillating at $e^{-i\omega_0 t}$ and $e^{-i(\omega_0+\Omega)t}$. Neglecting second-order time derivatives of the slow envelopes $a_{1,2}(t)$ and discarding non-resonant terms that couple to higher-order sidebands $e^{-i(\omega_0+n\Omega)t}$ with $|n|\ge2$ yields a pair of first-order differential equations for $a_{1,2}(t)$. The effective coupling strength $\kappa$ is proportional to the modulation depth $m$ and depends on the overlap between the unperturbed mode and the time-varying permittivity; explicit expressions can be obtained once a specific spatial mode profile is specified. The gain/loss rate $\gamma$ arises from a small imaginary component of the permittivity, which may itself be modulated in time to realize balanced gain and loss in the two Floquet components. The resulting equations are of the form of Eqs.~\eqref{eq:cmt1} and \eqref{eq:cmt2}, and the validity of the two-mode truncation requires $m\ll1$ and that the detunings to higher-order sidebands exceed the effective coupling strengths.

\begin{figure}[h]
\centering
\includegraphics[width=0.95\linewidth]{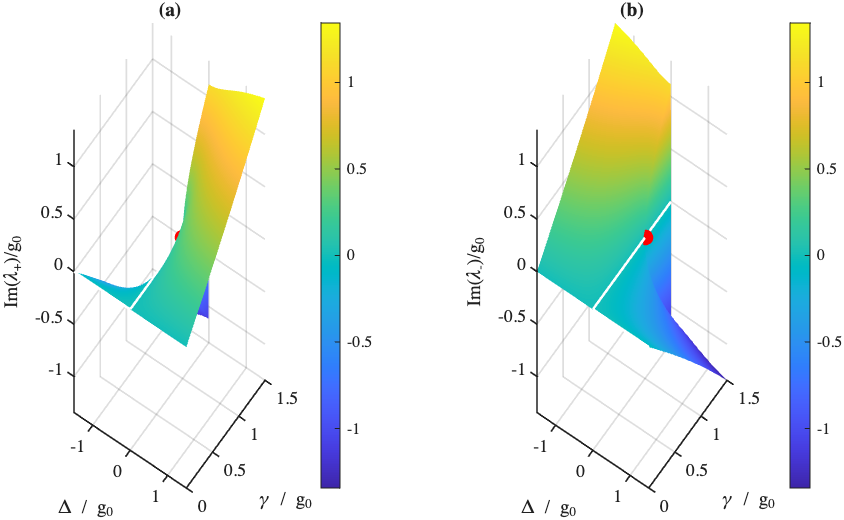}%{FigS1_Riemann_Im_exact.pdf}
\caption{Imaginary parts of the eigenvalue sheets $\lambda_{\pm}$ as functions of $(\Delta/g_0,\gamma/g_0)$.
The white contour indicates $\mathrm{Im}\,\lambda=0$, and the EP is located at $(\Delta,\gamma)=(0,\kappa)$.}
\label{fig:S1}
\end{figure}

Figure~\ref{fig:S1} plots the imaginary parts of the eigenvalues $\lambda_{\pm}$ over the $(\Delta/g_0,\gamma/g_0)$ plane, complementing the real-part surfaces shown in Fig.~\ref{fig:B}. The white contour marks the locus $\mathrm{Im}\,\lambda=0$, separating decaying and amplifying regimes. The EP lies at the intersection of the two sheets along this boundary.

\subsection*{S2. Berry phase versus loop radius}

\begin{figure}[h]
\centering
\includegraphics[width=0.75\linewidth]{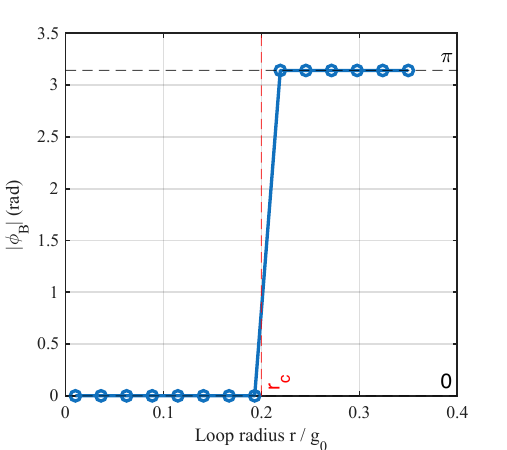}
\caption{Magnitude of the biorthogonal Berry phase $|\phi_B|$ as a function of the loop radius $r/g_0$ in the $(\Delta,\gamma)$ plane. A transition from $0$ to $\pi$ occurs when the loop begins to enclose the EP, demonstrating topological robustness.}
\label{fig:S2}
\end{figure}

In Fig.~\ref{fig:S2}, we examine how the magnitude of the biorthogonal Berry phase depends on the radius of the loop encircling the EP in the $(\Delta,\gamma)$ plane. For loops that do not enclose the EP, the accumulated phase is zero (mod $2\pi$). Once the loop radius exceeds a critical value $r_c$ such that the EP lies inside the loop, the Berry phase jumps to $\pi$ in magnitude and remains robust for larger radii. This step-like behavior confirms the topological nature of the EP.

\subsection*{S3. Exceptional-point coordinates versus coupling strength}

\begin{figure}[h]
\centering
\includegraphics[width=0.6\linewidth]{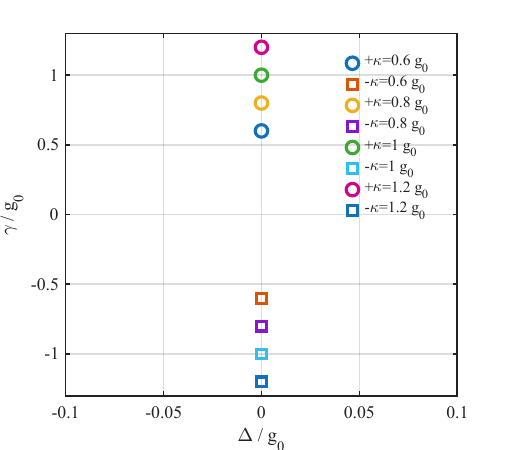}
\caption{Exceptional-point locations in the $(\Delta/g_0,\gamma/g_0)$ plane for several values of the coupling strength $\kappa/g_0$. All EPs occur at $\Delta=0$ and $\gamma=\pm\kappa$, confirming the analytic condition.}
\label{fig:S3}
\end{figure}

Figure~\ref{fig:S3} verifies the analytic EP condition in Eq.~\eqref{eq:EP_condition} by plotting the EP coordinates for several coupling strengths $\kappa$. In each case, the EP is found at $(\Delta,\gamma)=(0,\pm\kappa)$, in agreement with theory.

\subsection*{S4. Cram\'er--Rao bound versus Monte Carlo reconstruction}

\begin{figure}[h]
\centering
\includegraphics[width=0.95\linewidth]{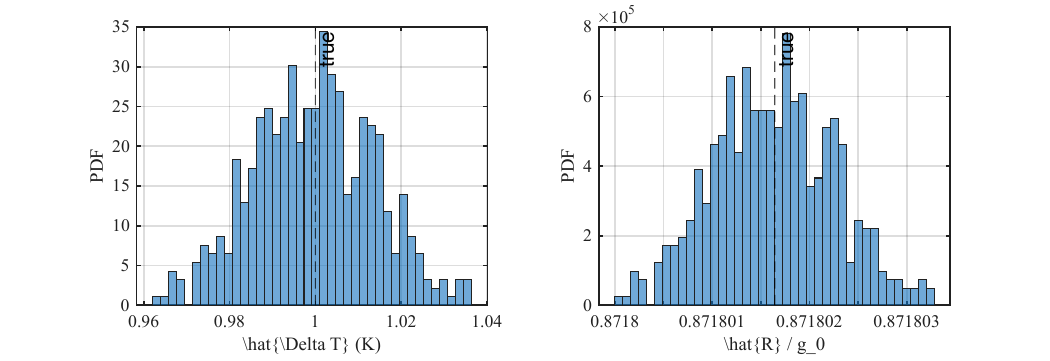}
\caption{Comparison of Cram\'er--Rao bounds (CRB) with Monte Carlo (MC) estimator statistics at a nonzero bias point.
Left: histogram of temperature estimates $\hat{\Delta T}$ with CRB indicated.
Right: histogram of splitting estimates $\hat{R}/g_0$, again in excellent agreement with the corresponding CRB.}
\label{fig:S4}
\end{figure}

Figure~\ref{fig:S4} compares the Cram\'er--Rao bound (CRB) for the temperature shift $\Delta T$ and the eigenvalue splitting $R$ to Monte Carlo (MC) histograms obtained by fitting noisy transmission spectra to the exact non-Hermitian model. The operating point is chosen such that $\partial R/\partial T\neq0$, ensuring a finite CRB for $R$. In both cases, the MC variances agree with the CRB predictions to within a few percent, confirming estimator optimality and validating the Fisher-information analysis used in the main text.

\subsection*{S5. Fit quality and splitting bias}

\begin{figure}[htb!]
\centering
\includegraphics[width=0.8\linewidth]{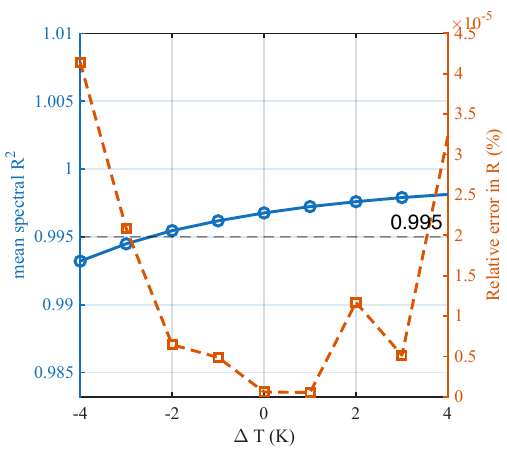}
\caption{Goodness-of-fit and splitting bias.
Left axis: mean spectral $R^2$ for fits to noisy transmission spectra as a function of $\Delta T$.
Right axis: relative error in the reconstructed splitting $R$ (in percent).}
\label{fig:S5}
\end{figure}

Finally, Fig.~\ref{fig:S5} quantifies the goodness-of-fit and relative bias in the reconstructed splitting across a range of temperature shifts. The mean spectral coefficient of determination $R^2$ exceeds $0.995$ for all $\Delta T$ in the range considered, and the relative error in $R$ remains below $\sim3\%$. These metrics confirm that the exact non-Hermitian transmission model provides an accurate and robust basis for parameter estimation in the proposed temporal EP sensing scheme.

\end{document}